\documentclass[aps,prl,twocolumn,superscriptaddress,showpacs]{revtex4-1}
\usepackage{}
\usepackage{txfonts}
\makeatletter

\newcommand{\Rmnum}[1]{\expandafter\@slowromancap\romannumeral #1@}
\usepackage{graphicx}
\usepackage{dcolumn}
\usepackage{bm}
\usepackage{epstopdf}
\usepackage{gensymb}
\usepackage{sidecap}
\usepackage{latexsym}
\usepackage{amssymb}
\usepackage{amsfonts}
\usepackage{soul}
\usepackage{color}
\usepackage{epstopdf}
\usepackage{placeins}
\usepackage{float}

\newcommand{\pcod}{Pr$_{2}$CuO$_{4\pm\delta}$\,}
\newcommand{\pcco}{Pr$_{2-x}$Ce$_{x}$CuO$_{4}$\,}
\newcommand{\rcco}{R$_{2-x}$Ce$_{x}$CuO$_{4}$\,}
\newcommand{\lcco}{La$_{2-x}$Ce$_{x}$CuO$_{4}$\,}
\newcommand{\plcco}{Pr$_{1-x}$LaCe$_{x}$CuO$_{4}$\,}

\begin{document}

\title{Normal State Gap in Parent Cuprate \pcod }

\author{Ge He}
\affiliation{Condensed Matter Physics, Institute of Physics, Chinese Acadamy of Sciences, Beijing 100190, China}

\author{Xinjian Wei}
\affiliation{Condensed Matter Physics, Institute of Physics, Chinese Acadamy of Sciences, Beijing 100190, China}

\author{Xu Zhang}
\affiliation{Condensed Matter Physics, Institute of Physics, Chinese Acadamy of Sciences, Beijing 100190, China}

\author{Lei Shan}
\affiliation{Condensed Matter Physics, Institute of Physics, Chinese Acadamy of Sciences, Beijing 100190, China}
\affiliation{Collaborative Innovation Center of Quantum Matter, Beijing 100190, China}
\affiliation{School of Physical Sciences, University of Chinese Academy of Sciences, Beijing 100190, China}

\author{Jie Yuan}
\affiliation{Condensed Matter Physics, Institute of Physics, Chinese Acadamy of Sciences, Beijing 100190, China}

\author{Beiyi Zhu}
\affiliation{Condensed Matter Physics, Institute of Physics, Chinese Acadamy of Sciences, Beijing 100190, China}

\author{Yuan Lin}
\affiliation{State Key Laboratory of Electronic Thin Films and Integrated Devices $\&$ Center for Information in Medicine, University of Electronic Science and Technology of China, Chengdu 610054, China}

\author{Kui Jin}\thanks{kuijin@iphy.ac.cn}
\affiliation{Condensed Matter Physics, Institute of Physics, Chinese Acadamy of Sciences, Beijing 100190, China}
\affiliation{Collaborative Innovation Center of Quantum Matter, Beijing 100190, China}
\affiliation{School of Physical Sciences, University of Chinese Academy of Sciences, Beijing 100190, China}

\date{\today}

\begin{abstract}
We present a tunneling study on single crystalline parent cuprate thin films, i.e. a series of \pcod (PCO) with tunable superconducting transition temperature. The zero-bias anomaly of differential conductance, well reported in the normal state of \rcco (R = Pr, Nd, La) and named as normal state gap (NSG), is observed in the Ce-free samples. This NSG behaves quite robust against the magnetic field up to 16 T, but fades away with increasing the temperature. Most importantly, we find that the magnitude of the NSG becomes larger with increasing point-contact junction resistance on the superconducting films, which is further enhanced in the non-superconducting samples of more oxygen disorders. The origination of NSG can be understood in the framework of Altshuler-Aronov-Lee (AAL) theory, where the disorder-induced electron-electron interactions suppress the density of states and thereby result in a soft Coulomb gap.
\end{abstract}

\pacs{74.72.Ek, 74.78.-w, 74.50.+r, 74.62.En}

\maketitle
The cuprates exhibit a great deal of anomalies besides the superconductivity, which are crucial to understanding the high $T_c$ mechanism, e.g. non-Fermi liquid behavior, pseudogap, etc \cite{Armitage2010,Scalapino2012,Keimer2015,Jin2011,Zhang2016}. The origination of the pseudogap is a protracted struggle for hole-doped cuprates, i.e. whether it is from phase incoherent Cooper pairs or other competing orders \cite{Ding1996,Renner1998,Norman2005,Ma2008,Kondo2009}. In electron-doped cuprates, there are two discriminable energy scales in the normal state, that is, the higher one (0.2$\sim$0.4 eV) mimicking a pseudogap and the so-called ``normal state gap" of lower energy (NSG, $\sim$5 meV) \cite{Armitage2010}. The former is observed by such as optical conductivity spectra \cite{Onose2001} and angle-resolved photoemission spectroscopy (ARPES) \cite{Armitage2002}, and identified as antiferromagnetic (AFM) spin correlations \cite{Onose2004}. Nevertheless, the NSG that usually behaves as a zero bias anomaly in differential conductance spectra remains controversial in its origin \cite{Biswas2001,Alff2003,Dagan2005,Shan2008}.

The zero bias anomaly in tunneling spectra may stem from various reasons, such as electron-electron interactions \cite{Altshuler1980}, Coulomb blockade \cite{Pekola2000}, hopping dominated conductance between the clusters of disordered metal grains \cite{Ossi2013}, Kondo scattering from magnetic moments, Giaever-Zeller two-step tunneling process, etc. In electron-doped cuprates, Alff et al. studied the tunneling spectra of \pcco and \lcco, and reported a $T^*$ that is smaller than $T_c$, pointing to a competing order below the superconducting dome \cite{Alff2003}. However, Dagan et al. reported that in \pcco the buildup temperature of NSG $T^*$ is slightly higher than $T_c$ in the underdoped region and approaching $T_c$ in the overdoped region, linked to the superconducting amplitude fluctuations \cite{Dagan2005}. By integrating the spectral weight and comparing the difference between the NSG and the superconducting state, Shan et al. provided an evidence to a two-gap scenario in \plcco \cite{Shan2008}. Such contradiction can be ascribed to the difficulty in defining the $T^*$, as well as the side effects from oxygen. It is known that slight oxygen variation is inevitable as tuning the Ce, which can result in remarkable influence on the physical properties \cite{Armitage2010}. As a special system of electron doped cuprates, the superconductivity of parent cuprates (i.e. R$_{2}$CuO$_{4\pm\delta}$) in $T'$ phase was discovered recently \cite{Matsumoto2008}. Most recently, optical conductivity measurements in \pcod (PCO) thin films disclosed that the high energy ``pseudogap" does not exist in this system \cite{Chanda2014}. However, the low energy NSG has never been addressed in this system, e.g. whether it is similar to other electron doped cuprates or not in such a Ce-free system.

In this work, we present a systemically tunneling study in PCO thin films with various $T_c$ by point-contact technique. The NSG is observed in this system, which is quite similar to other electron doped cuprates. The NSG is nearly field-independent but can be suppressed gradually with increasing the temperature. We find that there is a positive correlation between the magnitude of the NSG and junction resistance for all the superconducting samples, and the magnitude of the NSG is further enhanced in non-superconducting ones with more oxygen disorders. These phenomena can be well explained by AAL theory, revealing that the NSG stems from the disorder-induced electron-electron interactions.

\begin{figure*}[tb]
\includegraphics[width=1.5\columnwidth]{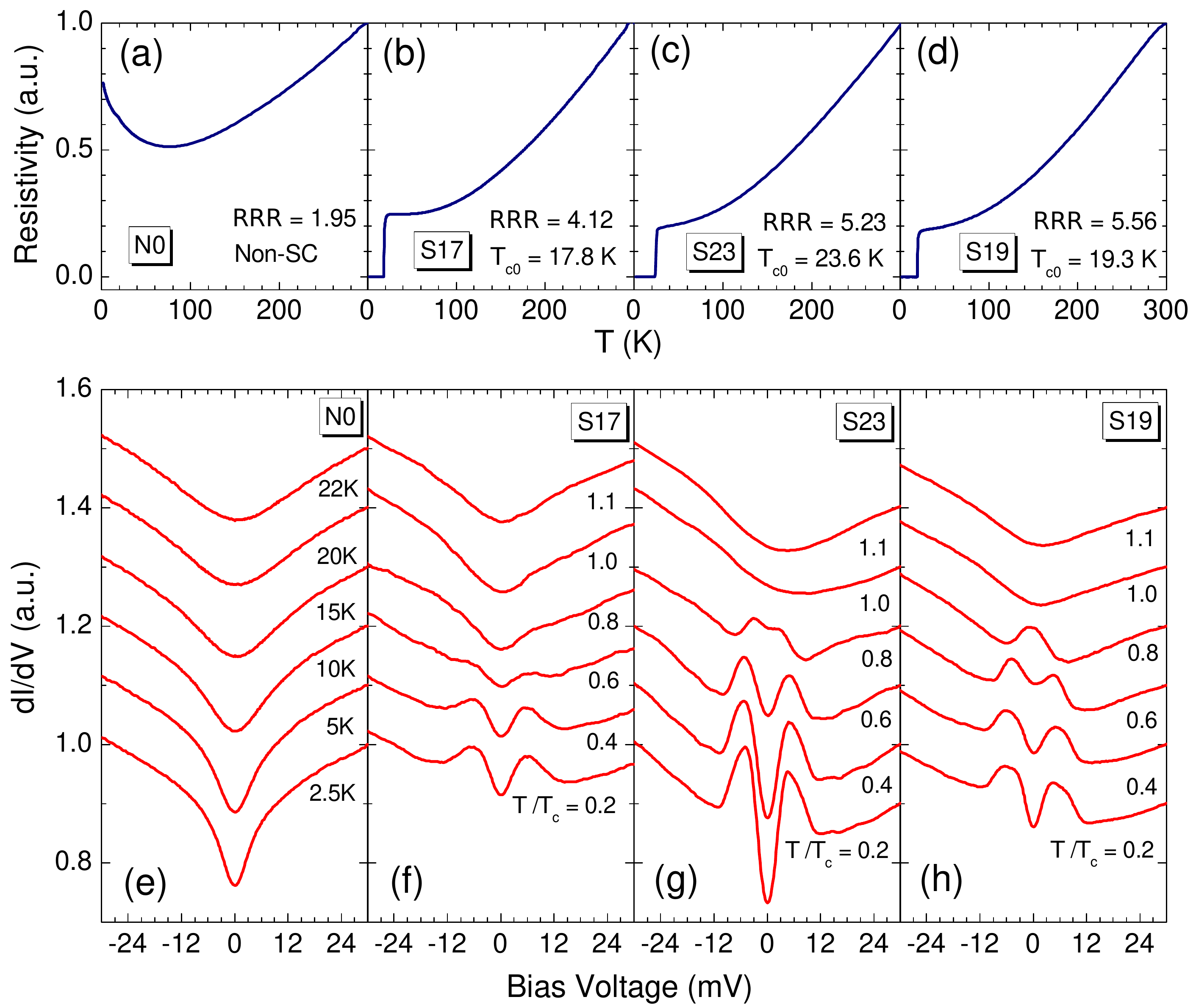}
\caption{(Color online) (a)~(d) Temperature dependence of resistivity for parent cuprate \pcod thin films with various $T_c$ and $RRR$. Resistivities are normalized by dividing the value at 300\,K. (e)~(h) $dI/dV$ versus bias voltage in various temperatures for these samples. All the curves are offset upwards for clarity.} \label{fig:sample}
\end{figure*}
The PCO thin films are grown by polymer assisted deposition (PAD) \cite{Jia2004,Lin2004} on (00l)-oriented SrTiO$_{3}$ substrate \cite{Wei2016}. The as-grown samples are fired at 850 $\textordmasculine$C in sealed tube with oxygen pressure at 200 Pa for crystallization. Then these samples are annealed at 400-600 $\textordmasculine$C under oxygen pressure of ~15 Pa. By adjusting the anneal temperature and time, samples with various $T_c$ can be obtained. The ab-plane resistivity is measured from 2 to 300\,K by a standard four-probe method using Quantum Design PPMS-16 equipment. We have selected six samples with full transition temperature $T_{c0}$ = 0 (N0), 15.5 K (S15), 16.4 K (S16), 17.8 K (S17), 19.3 K (S19) and 23.6 K (S23). Except for the non-superconducting sample N0, the others show narrow transition widths of $\Delta T$ = 1$\sim$2 K in the following measurements.

Tunneling spectra measurements are performed by a home-made point-contact probe, which can be put into the PPMS to ensure the temperature down to 2 K and field up to 16 T. Pt/Ir tips are used to make steady point-contact junctions. We measure the differential conductance spectra with a traditional lock-in technique. The spectra have good reproducibility for the same sample in various locations on the surface. The field is perpendicular to the ab-plane of the samples in all the measurements.
Figures 1(a)-1(d) show the temperature dependence of resistivity for N0, S17, S23 and S19. The residual resistance ratio ($RRR$) in non-superconducting sample is smaller than superconducting ones. Since the $RRR$ is sensitive to the amount of impurities and crystallographic defects, there should exist more disorders in non-superconducting sample. In electron doped cuprates, these disorders mainly come from the apical oxygen and in-plane oxygen vacancies induced by under- or over-annealing process \cite{Radaelli1994}. Figures 1(e)-1(h) present the $dI/dV$ versus bias voltage at various temperatures for above samples, respectively. Zero-bias anomaly observed in N0 demonstrates that the NSG state exists in the non-superconducting samples under zero field (see Fig.1(e)), similar to the non-superconducting Pr$_{1.89}$Ce$_{0.11}$CuO$_{4}$ sample \cite{Dagan2005}. Superconducting coherence peaks are observed in all the superconducting samples, which are suppressed with increasing temperature and disappear at $T_c$. The zero bias conductance is different among these samples due to the various effective barrier heights \cite{ Blonder1982}.

Figure 2(a) displays the spectra for S15 at both $T$ = 20 K, $H$ = 0 T and $T$ = 2.5 K, $H$ = 16 T. The spectra coincide with each other at bias higher than 7 mV, whereas the NSG state appears near the zero bias when field is applied to suppress superconductivity (see Fig. 2(a)). The spectra are almost unchanged with increasing field at $T$ = 2.5 K in N0 as seen in Fig. 2(b). Similar to the non-superconducting sample, the spectra are nearly the same in field up to 16 T after the coherence peaks are suppressed at $H$$\sim$6 T in S23 (see Fig. 2(c)), which is nearly consistent with the $H_{c2}$ measured in Ref. \cite{Krockenberger2012}. We define $G(30mV)/G(0)$ as the magnitude of NSG state and plot it as a function of $H$ as shown in Fig. 2(d). It can be clearly seen that the NSG is hard to be suppressed for all the samples even at $T$ = 15 K and $H$ = 16 T.

\begin{figure}[tb]
\includegraphics[width=1.0\columnwidth]{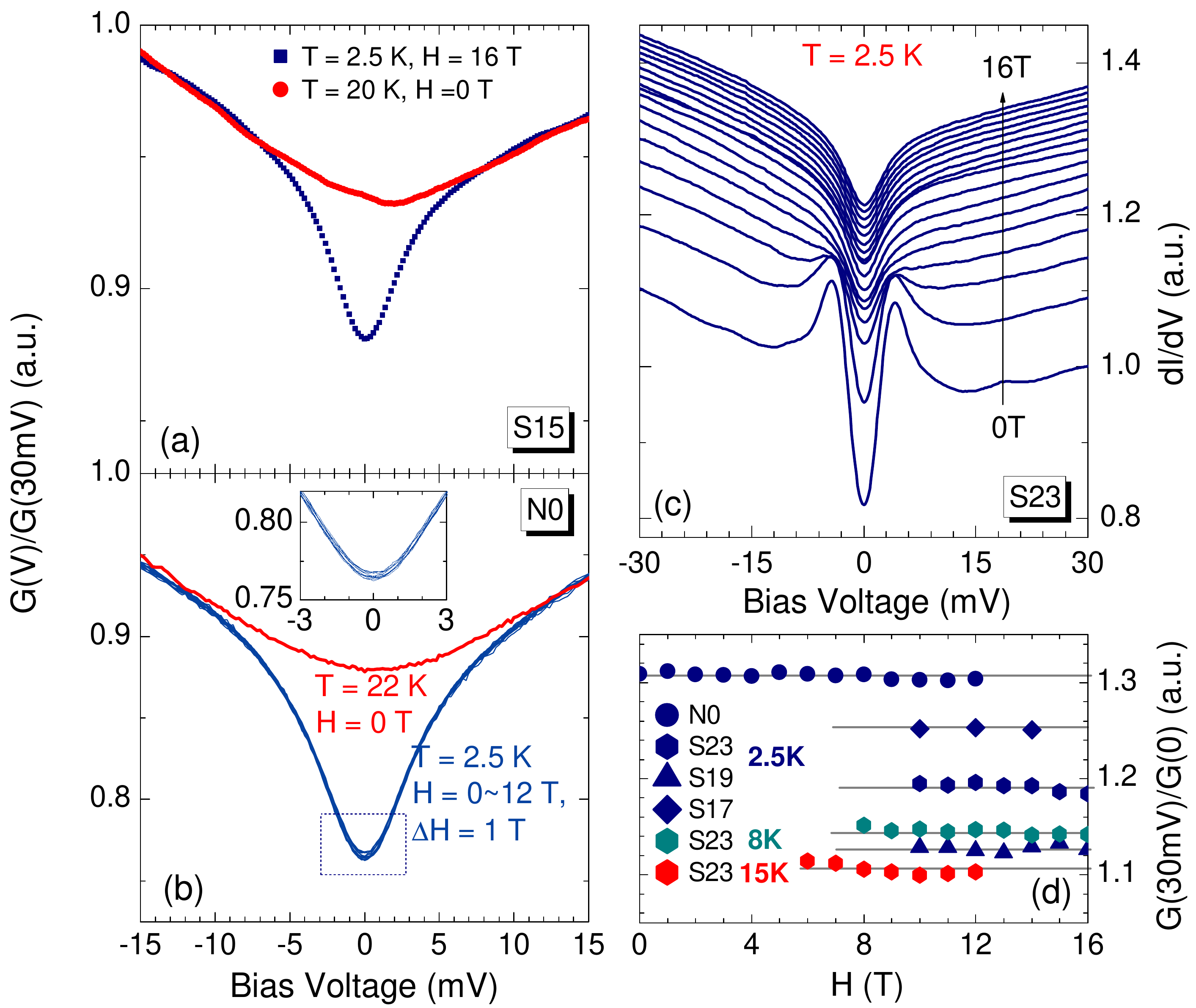}
\caption{(Color online) (a) $dI/dV$ versus bias voltage for S15 measured at temperature higher than $T_c$ (solid squares) with $H$ = 0 and magnetic field higher than $H_{c2}$ at $T$ = 2.5 K (solid circles). (b) $dI/dV$ of non-superconducting sample at $T$ = 2.5 K, $H$ = 0$\sim$12 T with $\Delta H$ = 1 T and $T$ = 22 K, $H$ = 0 T. Inset: zoom in the spectra near the zero bias (dashed square region). (c) $dI/dV$ versus bias voltage in different fields for S23. All curves are offset upwards for clarity. (d) Field dependence of $G(30mV)/G(0)$ for different samples. The horizontal gray lines is used for guiding eyes. For superconducting samples, the data are only plotted with fields higher than $H_{c2}$.} \label{fig:sample}
\end{figure}

As shown in Fig. 3(a), for the fields higher than $H_{c2}$, the zero bias dip in the spectra is continuously filled as increasing the temperature. Also, the $G(30mV)/G(0)$ decreases gradually with increasing the temperature (see Fig. 3(b)), which is quite similar to that in other electron doped cuprates \cite{Biswas2001,Dagan2005,Shan2008}. Taking into account the temperature induced Fermi function broadening effects, we calculate temperature dependence of the density of state based on the formula $N(eV, T) = \int{N(E,0)\frac{\partial f(E-eV,T)}{\partial E}dE}$. The calculated $G(30mV)/G(0)$ is obviously higher than experiments (see Fig. 3(b)), which suggests that Fermi broadening is not the main reason to close the NSG.

\begin{figure}[tb]
\includegraphics[width=1.0\columnwidth]{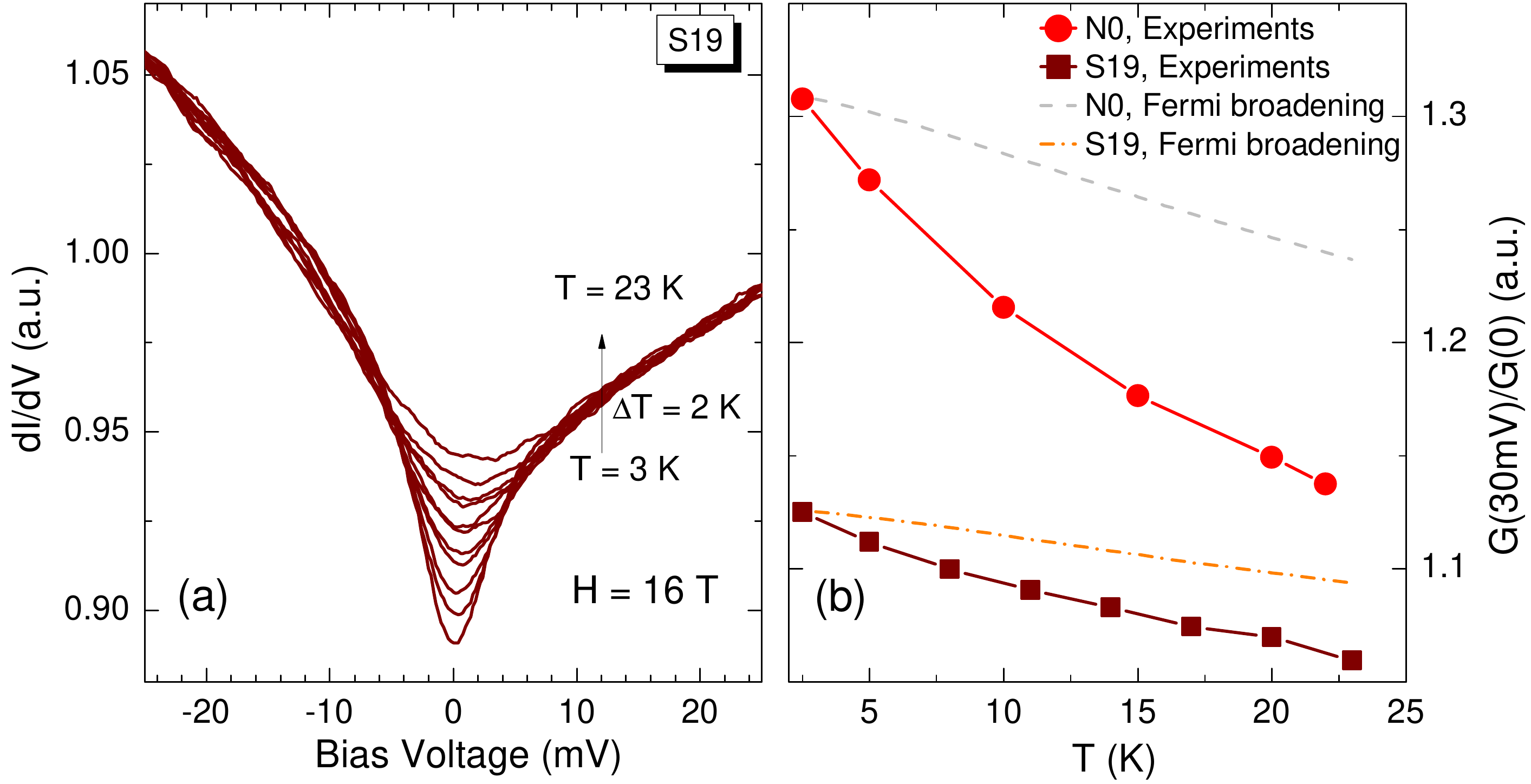}
\caption{(Color online) (a) $dI/dV$ for S19 in different temperatures at $H$ = 16 T. (b) Temperature dependence of $G(30mV)/G(0)$ for S19 (solid squares) at $H$ = 16 T and N0 at $H$ = 0 T (solid circles). The calculated $G(30mV)/G(0)$ with considering Fermi broadening effects for S19 (dash dotted line) and N0 (dashed line) are plotted as a function of temperature.} \label{fig:sample}
\end{figure}

In order to get further insight into the NSG state, we measure the $dI/dV$ spectra at various junction resistances ($R_j$) (see Fig. 4(a)). We find that the zero bias dip becomes deeper and deeper as increasing the $R_j$. The magnitude of NSG versus $R_j$ is plotted in Fig. 4(b), which shows a nearly positive relationship with $R_j$ for all the superconducting samples. Moreover, the magnitude of NSG in N0 is further enhanced compared to the superconducting ones. According to the Blonder-Tinkham-Klapwijk (BTK) theory, the normalized $dI/dV$ in the superconducting state comes from the contributions of two processes, i.e. the tunneling process and the Andreev reflection process \cite{Blonder1982}. The effective barrier height ($Z$) involving the thickness of oxide barrier and the mismatch of Fermi velocity impacts the contribution fractions from the two processes. On one hand, reducing $Z$ benefits the process of Andreev reflection and suppresses the tunneling process. On the other hand, increasing the oxide barrier thickness will enhance the scattering ratio and shorten the lifetime of quasiparticles, so it will weaken the measurement signal \cite{Dynes1984}. In the normal state, only tunneling process contributes to $dI/dV$. The increased $R_j$ mainly results from the increasing of oxide barrier thickness and must weaken the magnitude of NSG. However, it is contrary to our results.

\begin{figure}[tb]
\includegraphics[width=1.0\columnwidth]{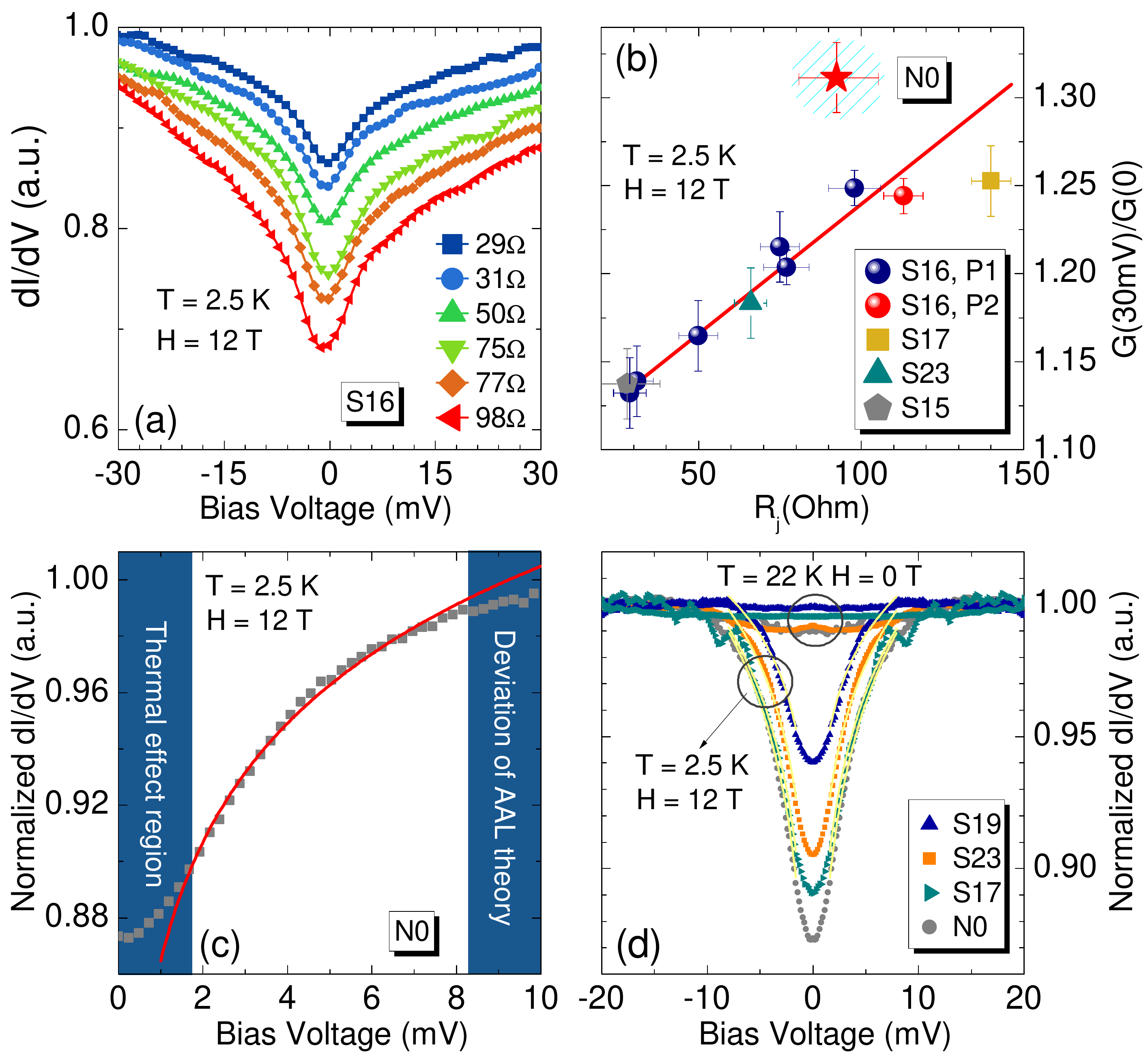}
\caption{(Color online) (a) $dI/dV$ versus bias voltage of S16 in different junction resistances at $T$ = 2.5 K and $H$ = 12 T. (b) $G(30mV)/G(0)$ versus junction resistance for different samples. The data of the superconducting samples can be nearly fitted with a solid line. The $G(30mV)/G(0)$ of non-superconducting sample (star) is higher than that of the superconducting samples at the same junction resistance. (c) Normalized $dI/dV$ versus bias voltage for N0 at $T$ = 2.5 K, $H$ = 12 T (solid squares). The data are fitted with AAL theory (solid lines). (d) Normalized $dI/dV$ versus bias voltage of different samples at $T$ = 2.5 K, $H$ = 12 T and $T$ = 22 K, $H$ = 0 T. All the data can be well fitted in the range of 2$\sim$8 mV (solid lines).} \label{fig:sample}
\end{figure}

We now summarize the feature of NSG in PCO: (1) NSG state is not sensitive to magnetic field in all the samples. (2) NSG can be suppressed easily by increasing temperature. (3) The magnitude of NSG is positively associated with $R_j$. (4) The magnitude of NSG from non-superconducting sample is further enhanced compared to the superconducting ones.

The Nernst behavior in Pr$_{2-x}$Ce$_{x}$CuO$_{4\pm\delta}$ discloses that the buildup temperature of superconducting fluctuations always follows the $T_c$ dome \cite{Li2007}. No matter for the superconducting fluctuations with the Maki-Thompson type or the Aslamazov-Larkin type above $T_c$, magnetic field should play a role in pair breaking or phase decoherence \cite{Larkin2005}, and therefore suppress the superconducting fluctuations. However, the NSG state persists in field up to 16 T, even in the non-superconducting sample. Besides, the magnitude of NSG is almost the same at $T$ = 15 K in field up to 12 T in PCO (see Fig. 2(d)). Moreover, we observe enhanced magnitude of NSG in non-superconducting sample. Thus, the superconducting fluctuations should not be the key reason for the NSG.

Based on our results, we argue that the NSG should stem from disorder-induced electron-electron interactions. Considering the interaction effects in disordered 2D Fermi systems, Altshuler et al. obtained a logarithmic-correction density of states. In tunneling experiments, the normalized corrections to the density of states can be given as \cite{Altshuler1980,Lee1985}:
\begin{equation}\label{1}
\frac{\delta N(\varepsilon)}{N_{1}} = \frac{1}{4\pi\varepsilon_{F}\tau} ln(2\kappa\Delta)ln(|\varepsilon|\tau),
\end{equation}
where $\delta N(\varepsilon)$ is the corrections to the density of states, $N_{1}$ is unperturbed density of states, $\varepsilon_{F}$ is the Fermi energy, $\tau$ is the relaxation time, $\Delta$ is the thickness of the barrier and $\kappa$ is the inverse screening length in 2D. Both $\tau$ and $\kappa$ can be used to describe the degree of disorders. Enhancing the degree of disorders will increase the corrections to the density of states. The enhanced $\Delta$ also leads to a larger $\delta N$. The AAL theory was confirmed by a number of tunneling experiments in various disordered metallic films, e.g. Be \cite{Butko2000}, Ag \cite{Yu2003} and In \cite{Yoshizawa2015}. We fit the normalized $dI/dV$ of various samples with AAL theory as seen in Figs. 4(c) and 4(d). All the data can be well fitted from 2 mV to 8 mV. The deviation at lower bias comes from the thermal broadening effect \cite{Biswas2001}. At high bias ($>$ 8 mV), the deviation results from other effects such as band edge effects \cite{Wang2012}, the break-down of WKB approximation \cite{Wei1998}, inelastic electron tunneling process \cite{Kirtley1990}, etc.

Now, we try to understand the behaviors of NSG state in the framework of AAL theory. A small amount of disorders in materials mainly cause two effects, i.e. weak localization and electron-electron correlations. The former stems from enhanced back scattering by quantum interference, which leads to a smaller conductance. The magnetic field destroys the quantum interference and increases the conductivity \cite{Hikami1980}. The latter originates from the destruction of long-range Coulomb screening. When the disorder-limited mean free path $l$ is reduced and comparable to the Fermi wavelength, i.e. $k_{F}l$ $\sim 1$, the density of states near the Fermi energy is obvious suppressed by the enhanced electron-electron Coulomb interactions \cite{Altshuler1980}, which is not sensitive to the magnetic field \cite{Lee1985}. The electron-electron interactions induce the localization of electrons as the electronic degrees of freedom freeze. With increasing the temperature, the localized electrons can gradually overcome the Coulomb interactions due to the thermal excitations, and the reduced density of states rebuilds. In addition, $R_j$ is adjusted mainly by changing the thickness of the oxide barrier. The larger thickness of oxide barrier leads to stronger corrections to the density of states \cite{Lee1985}. As mentioned above, the degree of disorder in N0 is stronger than that in the superconducting samples, which leads to an enhancement in magnitude of NSG.

In conclusion, we observe the NSG state in the Ce-free PCO thin films with tunable $T_c$. The NSG exhibits field-independence but temperature-dependence for both superconducting and non-superconducting samples. Importantly, there is a positive correlation between the magnitude of NSG and the junction resistance, and the magnitude of NSG is further enhanced in non-superconducting samples. All these behaviors are well consistent with AAL theory, indicating that the NSG in electron doped cuprates stems from disorder-induced electron-electron correlations.

This work was supported by the National Key Basic Research Program of China (Grants No. 2015CB921000 and 2016YFA0300301), the National Natural Science Foundation of China (Grants No. 11674374 and 11474338), the Key Research Program of Frontier Sciences, CAS (Grant No. QYZDB-SSW-SLH008) and the Strategic Priority Research Program of the CAS (Grants No. XDB07020100 and XDB07030200), the Beijing Municipal Science and Technology Project (Grant No. Z161100002116011).

\end{document}